\documentclass[aps,reprint,amsmath,amssymb,showpacs,floatfix,superscriptaddress]{revtex4-1}

\usepackage[utf8]{inputenc}
\usepackage[colorinlistoftodos]{todonotes}
\usepackage[colorlinks,citecolor=blue,linkcolor=blue]{hyperref}
\usepackage{graphicx}	
\usepackage{dcolumn}	
\usepackage{bm}			
\usepackage{txfonts} 	
\usepackage{siunitx}    

\newcommand{\CNN}{Centre de Nanosciences et de Nanotechnologies, CNRS, Univ. Paris-Sud, Universit{\'e} Paris-Saclay, 91120 Palaiseau, France}
\newcommand{\LMOPS}{LMOPS EA 4423 Lab, CentraleSup\'elec, Universit\'e Paris-Saclay and Universit\'e de Lorraine, 57070 Metz, France}
\newcommand{\Chaire}{Chaire Photonique, CentraleSup\'{e}lec, Universit\'{e} Paris-Saclay, 57070 Metz, France}
\newcommand{\IJL}{Institut Jean Lamour, CNRS, Universit\'{e} de Lorraine, 54011 Nancy, France}
\newcommand{\UMPhy}{Unit{\'e} Mixte de Physique, CNRS, Thales, Univ. Paris-Sud, Universit{\'e} Paris-Saclay, 91767 Palaiseau, France}

\begin{document}

\title{Pattern generation and symbolic dynamics in a nanocontact vortex oscillator}

\author{Myoung-Woo Yoo}
\email{myoung-woo.yoo@c2n.upsaclay.fr}
\affiliation{\CNN}
\author{Damien Rontani}
\affiliation{\LMOPS}
\affiliation{\Chaire}
\author{J{\'e}r{\'e}my L{\'e}tang}
\affiliation{\CNN}
\author{S{\'e}bastien Petit-Watelot}
\affiliation{\IJL}
\author{Thibaut Devolder}
\affiliation{\CNN}
\author{Marc Sciamanna}
\affiliation{\LMOPS}
\affiliation{\Chaire}
\author{Karim Bouzehouane}
\affiliation{\UMPhy}
\author{Vincent Cros}
\affiliation{\UMPhy}
\author{Joo-Von Kim}
\email{joo-von.kim@c2n.upsaclay.fr}
\affiliation{\CNN}

\date{\today}

\renewcommand{\figurename}{Figure}
\makeatletter
\renewcommand*{\fnum@figure}{{\normalfont\bfseries \figurename~\thefigure}}
\renewcommand*{\@caption@fignum@sep}{\textbf{ $|$ }}
\makeatother

\begin{abstract}
Harnessing chaos or intrinsic nonlinear behaviours from dynamical systems is a promising avenue for the development of unconventional information processing technologies. However, the exploitation of such features in spintronic devices has not been attempted despite the many theoretical and experimental evidence of nonlinear behaviour of the magnetization dynamics in nanomagnetic systems. Here, we propose a first step in that direction by unveiling and characterizing the patterns and symbolic dynamics originating from the nonlinear chaotic time-resolved electrical signals generated experimentally by a nanocontact vortex oscillator (NCVO). We use advanced filtering methods to dissociate nonlinear deterministic patterns from thermal fluctuations and show that the emergence of chaos results in the unpredictable alternation of simple oscillatory patterns controlled by the NCVO's core-polarity switching. With phase-space reconstruction techniques, we perform a symbolic analysis of the time series to assess the level of complexity and entropy generated in the chaotic regime. We find that at the centre of its incommensurate region, it can exhibit maximal entropy and complexity. This suggests that NCVOs are promising nonlinear nanoscale source of entropy that could be harnessed for information processing.
\end{abstract}

\maketitle

Nonlinear dynamics and chaos are powerful frameworks with which many phenomena in physics, biology, and engineering can be understood~\cite{Strogatz:2015}. Chaos refers to the high sensitivity of a nonlinear dynamical system to perturbations in its initial conditions, where the temporal evolution is unpredictable on the long term. From the perspective of applications in information processing, chaos has attracted much attention over the last two decades~\cite{Sciamanna:2015ec} because chaotic waveforms are random-like, yet deterministic and potentially controllable. They have found various applications in information technologies, such as encrypted communications at the physical layer~\cite{Argyris:2005br}, ultrafast random number generation~\cite{Uchida:08,Li:2013il}, data processing~\cite{Rontani:16}, computing~\cite{Ditto:2015fx}, secure-key exchange~\cite{Fischer:17}, radar applications~\cite{Liu:04}, precision sensing~\cite{Corron:01}, and encoding information via symbolic dynamics~\cite{Ott:93}.

Spintronic devices based on magnetic multilayers are good candidates for chaos-based applications, because magnetization dynamics in ferromagnets is intrinsically nonlinear~\cite{Wigen:1994wp, Alvarez:2000gt, Bertotti:2001ab, Lee:2004jo, Yang:2007ey, Slavin:2009fx, Pylypovskyi:2013ce, Bondarenko:2019fm, Montoya:2019fj}. Moreover, such dynamics can be driven and detected by spin-dependent transport phenomena, such as spin-transfer torques, magnetoresistive effects, and (inverse) spin Hall effects~\cite{Ralph:2008kj, Hoffmann:2013el}, giving rise to devices such as spin-torque and spin-Hall nano-oscillators~\cite{Kim:2012du, Chen:2016bf} that can be integrated into conventional semiconductor electronics~\cite{Villard:2010gi}. Because magnetization dynamics can occur at the nanoscale at microwave frequencies, spintronic devices hold much promise for highly-compact, GHz-rate information processing using chaos.

One example of chaos in a nanoscale spintronic device can be found in nanocontact vortex oscillators (NCVOs)~\cite{PetitWatelot:2012be, Devolder:2019uo}. In the NCVO, the gyration and switching of the vortex core can be induced by spin-transfer torques and oscillating output signals can be detected by the magnetoresistance. In contrast to vortex oscillators based on nanopillars~\cite{Pribiag:2007dk, Dussaux:2010ef, Locatelli:2011hw}, the NCVO can exhibit nontrivial dynamics that involves a self phase-locking phenomenon between the core gyration and core switching~\cite{PetitWatelot:2012be}. If the ratio between the frequencies of these two processes is irrational, the behaviour is chaotic~\cite{Devolder:2019uo}.

Here, we demonstrate experimentally that the chaotic regime of the NCVO involves simple aperiodic waveform patterns. These can be encoded into bit sequences, which are correlated with the core-polarity state of the magnetic vortex. First, we describe time-resolved signals from the NCVO at 77 K and validate their chaotic characteristics from sensitivity to initial conditions and correlation dimension analysis. Then, we show that the time traces are in fact only composed by a few waveform patterns which are ordered aperiodically in the chaotic regime. By reconstructing attractor geometries from the measured time series, we reveal the symbolic dynamics of chaotic NCVOs, which is in good agreement with the patterns observed in simulation. We extract bit sequences based on this symbolic analysis and show that the generated bits can achieve maximal values of the Shannon block entropy and Lempel-Ziv complexity.

%
\begin{figure*}[t!]
\centering
\includegraphics[width = 14cm]{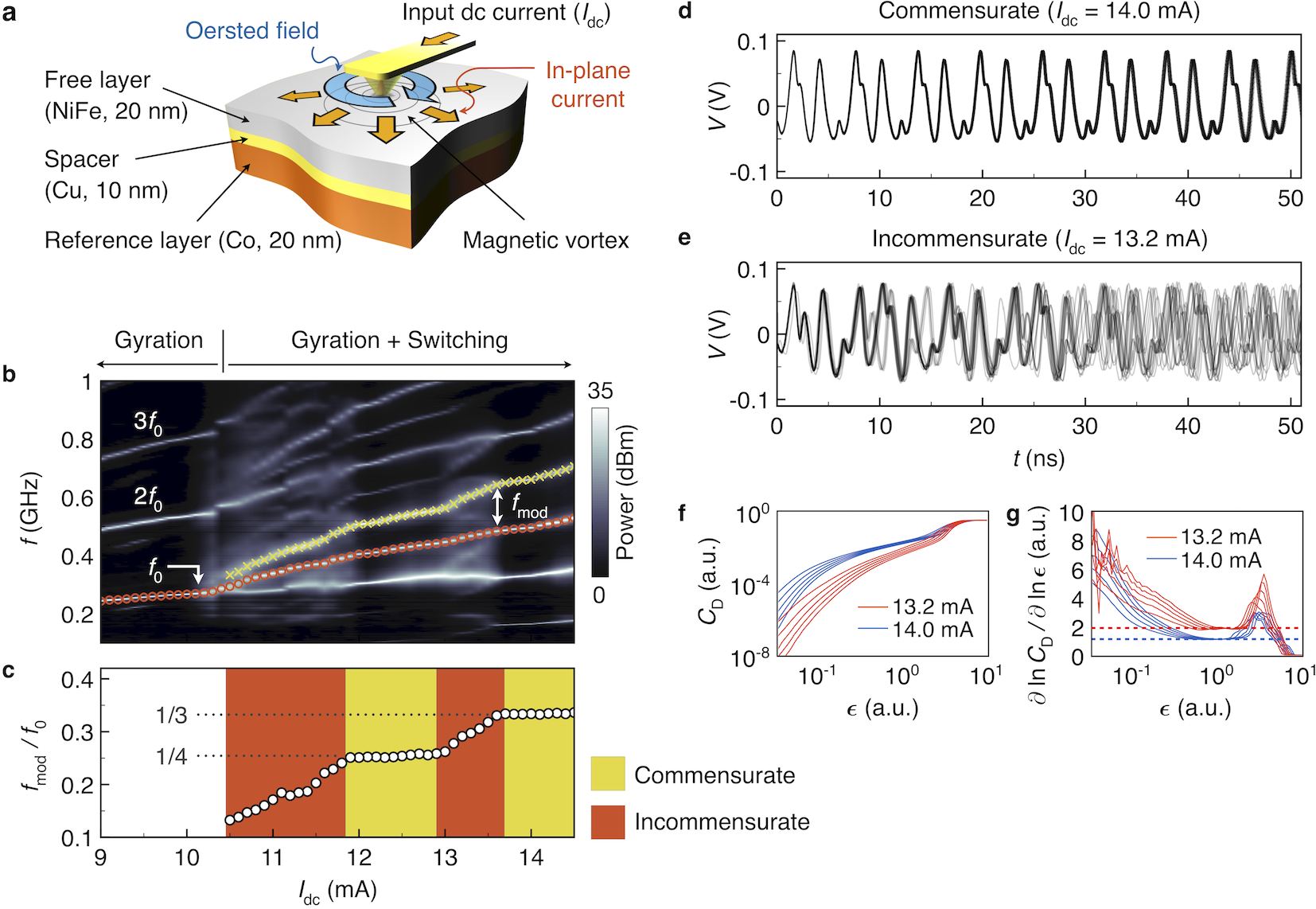}
\caption{\label{Fig:SpectraMap}
\textbf{Chaotic characteristics of the output time traces}. \textbf{a}, A schematic of a nanocontact vortex oscillator. \textbf{b}, A map of the power spectra as a function of input current amplitudes, $I_\mathrm{dc}$. The red circles and yellow cross marks indicate a fundamental frequency, $f_0$, and its upper sideband at $f_0 + f_\mathrm{mod}$, respectively. $f_\mathrm{mod}$ is a modulation frequency. \textbf{c}, $f_\mathrm{mod} / f_0$ as a function of $I_\mathrm{dc}$. The yellow and red regions represent the commensurate and the incommensurate states, respectively. The dotted horizontal lines indicate plateaus in which the self-phase-locking occurs. \textbf{d}, Eighteen different time traces which have identical initial conditions in the commensurate state ($I_\mathrm{dc}$ = 14.0 mA). \textbf{e}, as in \textbf{d} but in the incommensurate state at $I_\mathrm{dc}$ = 13.2 mA. \textbf{f} Correlation dimension, $C_\mathrm{D}$ as a function of a geometric scaling $\epsilon$ at $I_\mathrm{dc}$ = 13.2 mA (red lines) and 14.0 mA (blue lines). \textbf{g}, The derivatives of $C_\mathrm{D}$, $\partial \ln C_\mathrm{D} / \partial \ln \epsilon$. The red and blue dashed lines indicate the proper correlation dimensions at the flats for $I_\mathrm{dc}$ = 13.2 mA and 14.0 mA, respectively.
}
\end{figure*}

The NCVO comprises an extended spin-valve multilayer with a metallic point contact (approximately 20 nm in diameter) on the top of the surface (Fig.~\ref{Fig:SpectraMap}a)~\cite{PetitWatelot:2012be}. When an electric current is applied through the contact, the component of the current flow perpendicular to the film generates an Oersted field (blue arrow in Fig.~\ref{Fig:SpectraMap}a) which promotes a magnetic vortex in the free layer and generates a Zeeman energy potential for it that is centred on the nanocontact~\cite{Mistral:2008js}. The current component in the film plane (orange arrows in Fig.~\ref{Fig:SpectraMap}a) pushes the vortex core out from the centre by exerting spin-transfer torques~\cite{Zhang:2004hs}. The competition between the two effects results in a stable gyration~\cite{Pufall:2007jc, Mistral:2008js, PetitWatelot:2012be, Keatley:2016ku} and switching dynamics~\cite{PetitWatelot:2012be} of the vortex core around the nanocontact. In general, the shape of the core trajectory around the nanocontact is not circular~\cite{Keatley:2016ku}, as evidenced by a rich harmonic content in the power spectrum. This results from the presence of an antivortex or domain walls in the extended film that appear during the nucleation process~\cite{PetitWatelot:2012be}.

Fig.~\ref{Fig:SpectraMap}b shows a map of the power spectral density of the magnetoresistance oscillations at 77 K as a function of the applied current $I_\mathrm{dc}$, measured with a spectrum analyser. The NCVO exhibits three dynamical regimes: pure-gyration, commensurate, and incommensurate states~\cite{PetitWatelot:2012be}. When $I_\mathrm{dc}$ is lower than a threshold for core reversal ($\sim$10.3 mA here), only the gyration frequency, $f_0$, is observed (red circles in Fig.~\ref{Fig:SpectraMap}b) with its harmonics ($2 f_0$, $3 f_0$, \dots) because of the noncircular core trajectories. If $I_\mathrm{dc}$ is larger than the threshold, core reversal appears in addition to the gyration. This dynamical state is accompanied by additional sidebands at $f_0 \pm f_\mathrm{mod}$ (yellow crosses in Fig.~\ref{Fig:SpectraMap}b), where $f_\mathrm{mod}$ is a modulation frequency that is related to the periodicity of the core reversal. The ratio $f_\mathrm{mod} / f_0$ as a function of $I_\mathrm{dc}$ is shown in Fig.~\ref{Fig:SpectraMap}c, which is distinguished by two plateaus with monotonic increases elsewhere. At the plateaus (yellow regions in Fig.~\ref{Fig:SpectraMap}c), the frequency spectrum shows clear peaks and $f_\mathrm{mod} / f_0$ remains constant with $I_\mathrm{dc}$. In this case, the ratio can be expressed as simple integer fractions ($1/3$ and $1/4$ in this experiment), because the core reversal process is phase-locked to the gyration~\cite{PetitWatelot:2012be}. The relation between the core dynamics and $f_\mathrm{mod} / f_0$ is discussed in more detail in the Supplementary Note 1 and Supplementary Figure 1. In the incommensurate state (red regions in Fig.~\ref{Fig:SpectraMap}c), in contrast, the frequency spectrum becomes more complex, where $f_\mathrm{mod} / f_0$ varies with $I_\mathrm{dc}$ from one plateau to another, which indicates that no phase locking occurs between the gyration and core reversal.

The incommensurate state represents chaotic behaviour~\cite{Devolder:2019uo}. Sensitivity to initial conditions, a hallmark of chaos, can be seen in the time traces. These were obtained at 77 K with a single-shot oscilloscope, which were then filtered using a pattern matching technique to reduce the measurement noise (Supplementary Note 3 and Supplementary Figure 3). By overlaying several segments of the time traces with very similar initial conditions, we can obtain a visual measure of the sensitivity to initial conditions in the commensurate (Fig.~\ref{Fig:SpectraMap}d) and incommensurate states (Fig.~\ref{Fig:SpectraMap}e). In the commensurate state (Fig.~\ref{Fig:SpectraMap}d), the waveforms remain coherent over tens of nanoseconds, with evidence of jitter setting in at around 50 ns. In the incommensurate state, however, the coherence is lost below 10 ns (Fig.~\ref{Fig:SpectraMap}e), which is due to the sensitivity to initial conditions. We can further verify the presence of chaos by analysing the fractal geometry of the reconstructed attractor~\cite{Kantz:2004th}. To this end, we compute the correlation dimension $D_c$ from the filtered time series using the correlation sum $C(\epsilon,m)$ in Fig.~\ref{Fig:SpectraMap}f and its derivative with respect to a geometric scaling $\epsilon$ (see Methods). We estimate the geometric dimension in Fig.~\ref{Fig:SpectraMap}g by looking at constant values of the derivative. For the commensurate case (blue lines in Fig.~\ref{Fig:SpectraMap}g) the dimension is found to be $\sim$1.04, which is very close to 1 and consistent with limit-cycle dynamics of a NCVO presenting a small amount of jitter in the position of the core-polarity switching. For the incommensurate case (red lines in Fig.~\ref{Fig:SpectraMap}g), however, the dimension is found to be $\sim$1.85, which is consistent with a fractal geometry associated with temporal chaos. These results are consistent with a previous study using the titration of chaos with added noise~\cite{Poon:01} to identify the presence of chaos in the NCVO~\cite{Devolder:2019uo}. Note that the responses in the incommensurate state can be reproduced by micromagnetic simulations even at 0 K as shown previously~\cite{PetitWatelot:2012be}. We contend therefore that the measured chaotic characteristics are mostly deterministic, rather than stochastic as driven by thermal fluctuations.

\begin{figure*}[t!]
\centering
\includegraphics[width = 14cm]{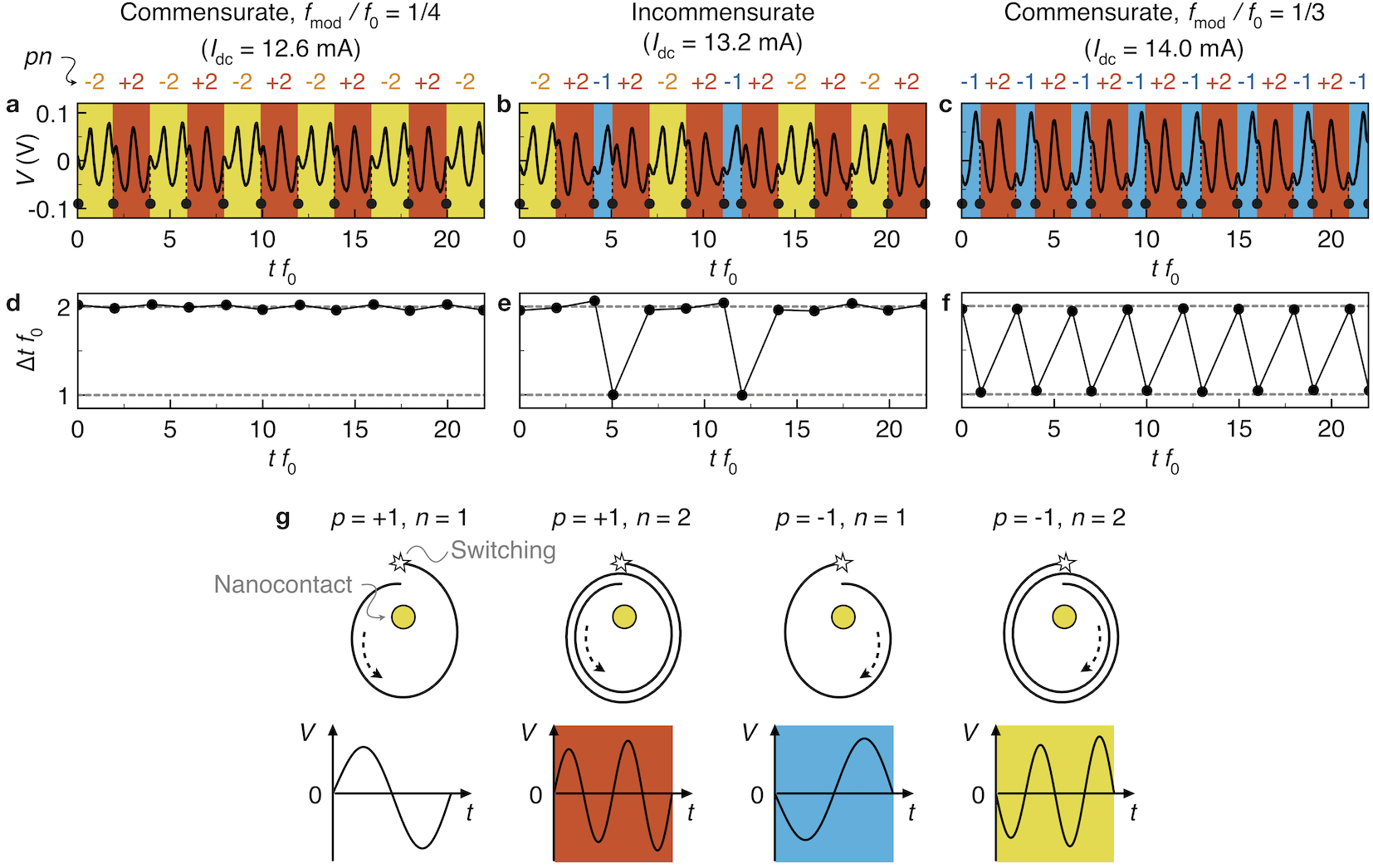}
\caption{\label{Fig:pn}
\textbf{Pattern generation from time series}. \textbf{a-c}, Representatives of experimentally measured time series at $I_\mathrm{dc}$ = 12.6 mA, 13.2 mA, and 14.0 mA. The black dots with dotted lines indicate the core-polarity switching events. We normalised the time axes $t f_0$ which is identical with a number of core gyrations. The yellow, blue, and red regions indicate waveform patterns denoted by $pn$ = $-2$, $-1$, and $+2$, respectively, where $p$ is a core polarity and $n$ is a required gyration number for the core-polarity switching. \textbf{d-f}, Time evolutions of a required gyration number for the core switching, $n = \Delta t f_0$, obtained from \textbf{a-c} by calculating intervals between the black dots. \textbf{g}, Schematics of the possible core-polarity switching scenarios. The core trajectories (top panels) and expected output waveforms (bottom panels) are shown for different $p$ and $n$ combinations. The colours of the oscillatory patterns (red, blue, and yellow) correspond with those in a-c.
}
\end{figure*}

A feature of the chaos generated by the NCVO involves distinct waveforms that repeat aperiodically. Representative experimental time series are shown in Fig.~\ref{Fig:pn}a-c. Here we use a time axis that is normalised with respect to the core gyration period, $1/f_0$, such that $t f_0$ represents the number of core gyrations. Both commensurate and incommensurate states show similar features where distinct oscillatory patterns are delimited by cusps. These patterns correspond to a number of orbits of the vortex core around the nanocontact, with the cusps representing a core reversal event; the position of these cusps are indicated by the dots and dotted lines in Fig.~\ref{Fig:pn}a-c. Note that core reversal results in the change in the sense of gyration (i.e. clockwise to counterwise, and vice versa). These features in the measured time series are reproduced in micromagnetic simulations (see Supplementary Note 2 and Supplementary Figure 2).

From the intervals between the core switching events in Fig.~\ref{Fig:pn}a-c, we can define a gyration number for the core switching, $n = \Delta t f_0$, which is shown in Fig.~\ref{Fig:pn}d-f. In the commensurate state, $n$ exhibits a simple time evolution. At $I_\mathrm{dc}$ = 12.6 mA, the switching always occurs every two gyrations (Fig.~\ref{Fig:pn}d), so $n$ remains constant at 2. Similarly, at $I_\mathrm{dc}$ = 14.0 mA, core reversal occurs after one and two gyrations successively, a process which repeats periodically; in this case $n$ oscillates between 1 and 2 as shown in Fig.~\ref{Fig:pn}f. In the incommensurate state ($I_\mathrm{dc}$ = 13.2 mA), however, $n$ switches between 1 and 2 in an aperiodic fashion (Fig.~\ref{Fig:pn}e), which is consistent with the chaotic dynamics expected in this regime.

The required gyration number for core switching, $n$, is always approximately integer (typically 1 or 2 in this experiment) in both commensurate and incommensurate cases as shown in Fig.~\ref{Fig:pn}d-f. This is consistent with simulation results, in which core-polarity switching occurs only in a restricted region of the film plane close to the nanocontact, where conditions for core reversal are met~\cite{PetitWatelot:2012be}. In addition the core polarity, $p$, can only have two  values, $+1$ and $-1$, so we hypothesize that in general there are only four possible patterns for the commensurate or incommensurate states, $pn$ = $+1$, $+2$, $-1$, and $-2$. In other words, the nonlinear physical properties of the NCVO result in a sequence that represents a combination of these four patterns.

We plot schematic core trajectories of the possible switching scenarios in Fig.~\ref{Fig:pn}g, along with the expected time series for different $pn$. Without loss of generality, we assume here that the vortex has clockwise chirality and the reference layer is saturated in the $+y$ direction. Based on the schematic waveforms, we can identify the corresponding oscillatory patterns from the time series, as indicated by the background colours in Fig.~\ref{Fig:pn}a-c. In the commensurate state (Fig.~\ref{Fig:pn}a and c), the time series are composed of two $pn$ patterns, which repeat periodically. They involve $pn$ = \{$-2$, $+2$\} for $I_\mathrm{dc}$ = 12.6 mA (Fig.~\ref{Fig:pn}a) and \{$-1$, $+2$\} for 14.0 mA (Fig.~\ref{Fig:pn}c), respectively. In the incommensurate state (Fig.~\ref{Fig:pn}b), there exist three $pn$ patterns, $pn$ = \{$-2$, $-1$, $+2$\}, which appear without a well-defined periodicity. This shows that the NCVO generates simple oscillatory patterns even in the chaotic state.

\begin{figure*}[t!]
\centering
\includegraphics[width = 14cm]{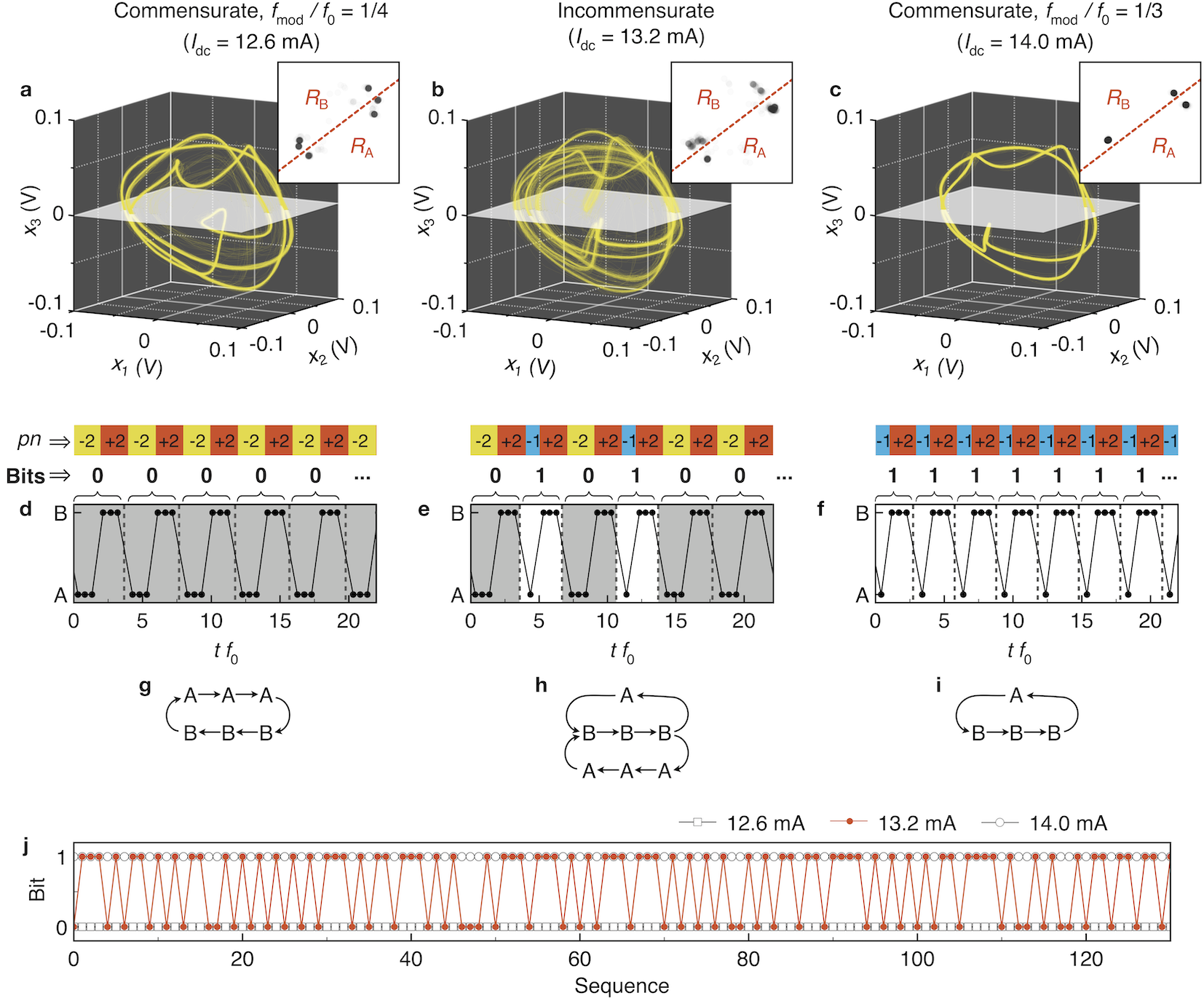}
\caption{\label{Fig:Symbolic}
\textbf{Reconstructed attractor geometries and symbolic dynamics}. \textbf{a-c}, Reconstructed attractor geometries by a method of delay from the measured time traces at $I_\mathrm{dc}$ = 12.6 mA, 13.2 mA, and 14.0 mA, respectively. The white plane is arbitrarily chosen Poincar{\'e} surface of section (see Methods). (Inset) Poincar{\'e} maps at the surfaces. The red dashed lines indicate a simple partition to divide the plane into two regions, $R_\mathrm{A}$ and $R_\mathrm{B}$ for encoding symbols, A and B. The Poincar{\'e} surface of section and partitions are identical for all $I_\mathrm{dc}$ in these figures. \textbf{d-f}, Dynamics of symbols defined from the partition on the Poincar{\'e'} maps. Above the graphs, corresponding $pn$ patterns and generated bit sequences are represented. The bits are defined as 0 $\equiv$ [A,A,A,B,B,B] and 1 $\equiv$ [A,B,B,B]. \textbf{g-i}, Rules of the symbolic dynamics at $I_\mathrm{dc}$ = 12.6 mA, 13.2 mA, and 14.0 mA. \textbf{j}, Generated bit sequences for long term at $I_\mathrm{dc}$ = 12.6 mA, 13.2 mA, and 14.0 mA.
}
\end{figure*}

We further analyse the pattern generation from the perspective of symbolic dynamics~\cite{Bollt:2003gm}. The principle of symbolic dynamics is to find an adequate partition of the system's Poincar\'{e} section in its phase space (see Methods), such that every time there is a transition from one region of the section to another, a symbol is emitted. As a result, the nonlinear dynamics of the system can be reduced to a sequence of symbols. However, finding the proper (said to be \textit{generating}) from model-free, experimental, scalar time series is a challenging problem in general. First, we reconstruct attractor geometries from the measured time series using a three-dimensional delay embedding, which is sufficiently large to completely unfold the reconstructed attractors when considering the commensurate and incommensurate  states shown in Fig.~\ref{Fig:Symbolic}a-c~\cite{Kantz:2004th}. In the commensurate state, the NCVO exhibits a limit cycle, because the trace is periodic (Fig.~\ref{Fig:Symbolic}a and c). On the contrary, the attractor for the incommensurate state is more intricate (Fig.~\ref{Fig:Symbolic}b). As explained in the Methods section, we set a proper Poincar{\'e} surface (white surfaces in Fig.~\ref{Fig:Symbolic}a-c) allowing for a potential generating partition. Note that the surface is not unique, and we can choose any different plane for the symbolic analysis. On the surface, we obtain the Poincar{\'e} maps from the intersection (insets of Fig.~\ref{Fig:Symbolic}a-c) in which we find several clusters of the points. Note that here, for the Poincar{\'e} map, we do not consider the transverse orientation. We use a simple partition to divide the map into two different regions, $R_\mathrm{A}$ and $R_\mathrm{B}$ (red dashed lines in the insets of Fig.~\ref{Fig:Symbolic}a-c), then record the symbols, A or B, when the attractor cross the surface either in $R_\mathrm{A}$ or $R_\mathrm{B}$, respectively. The encoded symbolic sequences are shown in Fig.~\ref{Fig:Symbolic}d-f, where the corresponding $pn$ patterns are shown above the graphs. By comparing the $pn$ patterns and symbolic dynamics, we can see that $pn$ = $-2$, $-1$, and $+2$ correspond to [A,A,A], [A], and [B,B,B], respectively. This result shows that the choice of the partition for the determination of the symbolic dynamics is in good agreement with the $pn$ sequences in both the commensurate and incommensurate states. Other partition choices are also possible but may  render identifying the symbolic sequences more difficult (see Supplementary Note 4 and Supplementary Figure 4).

We can find simple rules in the symbolic sequences (Fig.~\ref{Fig:Symbolic}d-f). In the commensurate case, the sequences show only one repeated cycle: [A,A,A,B,B,B] and [A,B,B,B] for $I_\mathrm{dc}$ = 12.6 mA and 14.0 mA, respectively (Fig.~\ref{Fig:Symbolic}g and i). In the incommensurate state, however, two possible cycles coexist in the sequence and appears erratically over time (Fig.~\ref{Fig:Symbolic}h). To simplify the analysis of complexity, we define binary symbols attributed to the two different patterns accessible by the NCVO: 0 $\equiv$ [A,A,A,B,B,B] and 1 $\equiv$ [A,B,B,B]. Then, we extract the bit sequences as represented in Fig.~\ref{Fig:pn}d-f above the graphs. In the commensurate states (Fig.~\ref{Fig:pn}d and \ref{Fig:pn}f), the NCVO generates only one type of bits: $0$ for $I_\mathrm{dc}$ = 12.6 mA and $1$ for $I_\mathrm{dc}$ = 14.0 mA. On the other hand, in the incommensurate state (Fig.~\ref{Fig:pn}e), the NCVO generates bits in no apparent order. To better illustrate this, we plot a bit sequence generated in the chaotic regime for a longer duration (Fig.~\ref{Fig:Symbolic}j). We note that the bits are generated at an average rate of $\sim$131 MHz, which is much faster than stochastic-based random number generators~\cite{Fukushima:2014kz, Vodenicarevic:2017db}. Interestingly, the possible cycles in the incommensurate state correspond to those the NCVO already exhibits in its commensurate state at $I_\mathrm{dc} = 12.6$ mA and $14.0$ mA, respectively (Fig.~\ref{Fig:Symbolic}h). This shows that not only the complexity is driven by the core switching, but also that the type of patterns generated in the incommensurate (chaotic) state is fundamentally restricted to accessible patterns associated to the two neighbouring commensurate states. Hence, within the incommensurate region, we anticipate that the probability of appearance of one among the two accessible patterns can be controlled by $I_\mathrm{dc}$.

\begin{figure}[t!]
\centering
\includegraphics[width = 8.5cm]{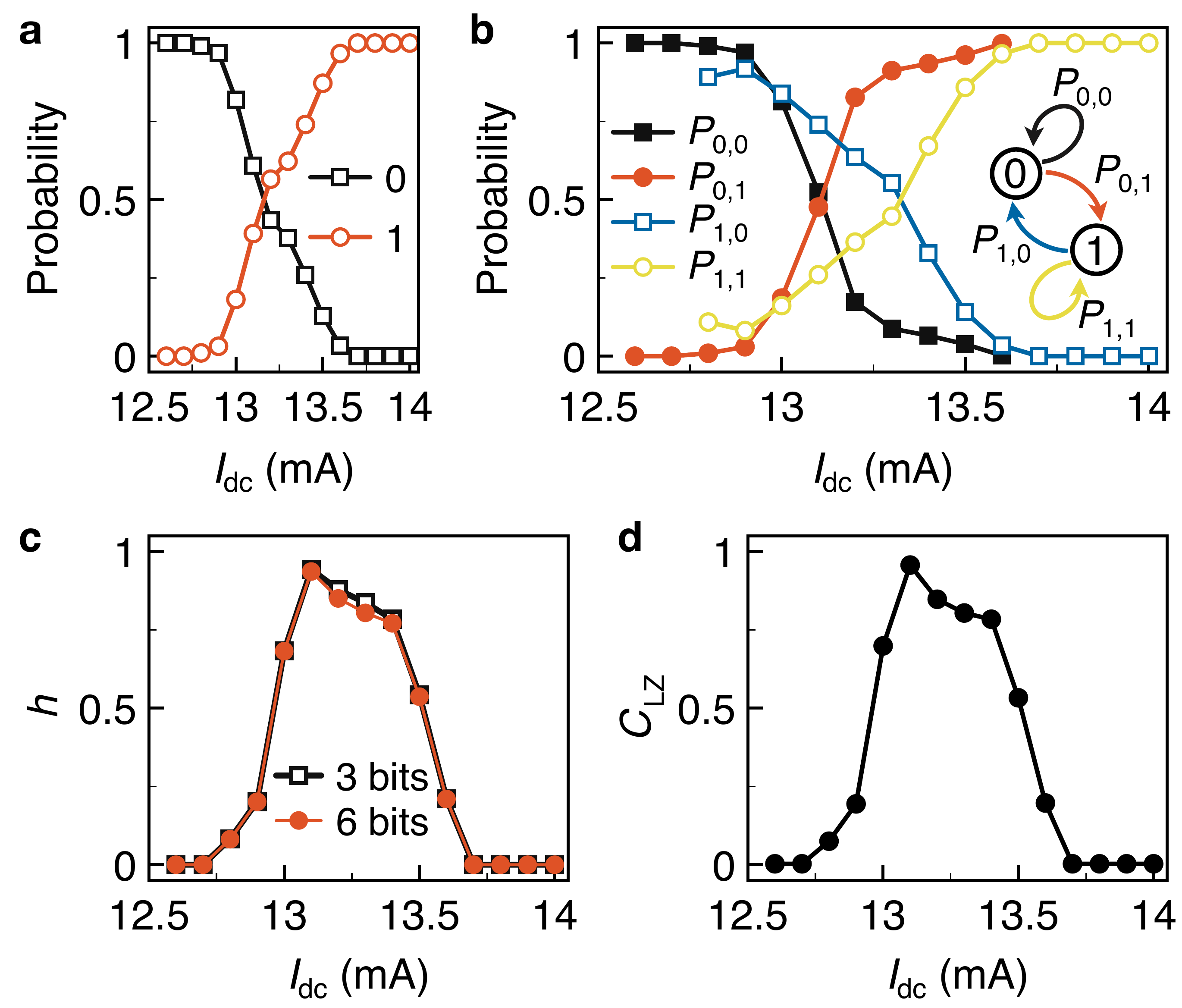}
\caption{\label{Fig:Complexity}
\textbf{Complexity and control of bit sequences}. \textbf{a}, Probability of 0 and 1 in the generated bit sequences as a function of $I_\mathrm{dc}$. \textbf{b}, Probability of moving from the current state, $i$, to the next state, $j$, $P_{i,j}$. The inset shows the Markov chain for a 1 bit information.\textbf{c}, Shannon block entropy rate, $h$, as a function of $I_\mathrm{dc}$ for different block lengths, 3 bits and 6 bits. \textbf{d}, Normalized Lempel-Ziv complexity, $C_\mathrm{LZ}$, as a function of $I_\mathrm{dc}$.
}
\end{figure}

To assess the complexity of the bit sequence extracted from the symbolic dynamics, we compute the probability of each bit as a function of $I_\mathrm{dc}$ (Fig.~\ref{Fig:Complexity}a). We also estimate the probability of transition from one bit to the next, $P_{i,j}$, where $i$ and $j$ are the current and the next binary states (Fig.~\ref{Fig:Complexity}b), while assuming a Markovian property, i.e., a one-bit memory depth associated to the bit sequence (see Supplementary Note 6 and Supplementary Figure 6). In Fig.~\ref{Fig:Complexity}a, as $I_\mathrm{dc}$ increases and the NCVO transitions from a commensurate to an incommensurate state, we observe that the probability of symbol 0 (resp. symbol 1) decreases smoothly from $P_0 = 1$ (resp. increases from $P_1 = 0$) monotonically and reaches the point where $P_0 = P_1 = 0.5$ at about $I_\mathrm{dc} = 13.1$ mA before continuing to decrease to $P_0 = 0$ (resp. to increase to $P_1 = 1$). Similarly, in Fig.~\ref{Fig:Complexity}b, we observe similar behaviour for the transition probabilities and specifically at $I_\mathrm{dc} \approx 13.1$ mA, we have $P_{0,0} = P_{1,1}$ and $P_{0,1} = P_{1,0}$. These results imply that in the incommensurate state, the symbolic dynamics of the NCVO is akin to a fair coin toss, where it becomes difficult to predict the next generated bit from the current one. But as the current is varied from this operating point, the NCVO ``coin'' becomes biased because the dynamics approaches more closely to one of the two neighbouring commensurate states. As such, one of the two accessible temporal patterns will start to dominate the other and its corresponding bit (0 or 1) will become more probable.

To further assess the unpredictability of the generated bits, we evaluate the complexity of the bit sequences generated by using two metrics from information theory: the Shannon block entropy rate, $h$, and the normalised Lempel-Ziv complexity, $C_\mathrm{LZ}$. The Shannon block entropy rate measures the amount of uncertainty carried on average by each bit: for a binary source of information, as the one generated by a chaotic NCVO, we have maximum entropy $h_\mathrm{max} = 1$ bit/binary symbol. We compute the entropy rate as a function of $I_\mathrm{dc}$ (Fig.~\ref{Fig:Complexity}c) and consider binary blocks of length $n = 3$ and $n = 6$ to obtain more robust estimates, while ensuring the estimation remains invariant with respect to these choices of block-lengths (see Methods). We observe that the uncertainty from the bit stream generated by the NCVO is non-monotonic inside the incommensurate region; it gradually increases to reach a peak value of $h \approx 0.94$ at $I_\mathrm{dc} = 13.1$ mA before decreasing again as the 1/4 commensurate state is approached. This is consistent with an asymmetric distribution for the probability mass function of the generated bit, which indicates that the bit sequence generated by the NCVO in its incommensurate state inherits complexity from the Kolmogorov-Sinai entropy created by the chaotic dynamics~\cite{Kantz:2004th}. We perform a similar analysis for the Lempel-Ziv complexity $C_\mathrm{LZ}$, which measures the diversity (\emph{i.e.}, lack of redundancy) of binary patterns encountered in a binary sequence. We observe a similar trend with finite values in the incommensurate region, reaching almost the maximum value of $C_\mathrm{LZ,max} = 1$, while $C_\mathrm{LZ} ~\simeq 0$ in the commensurate state.  This means the bits generated by the chaotic NCVO cannot be efficiently compressed because of its maximal complexity. Since $h$ and $C_\mathrm{LZ}$ almost attain their maximum values of $\sim$1, the raw generated bit sequences have suitable statistical features to be considered as a physical source of entropy for information processing. The probability and complexity assessments have been performed on more than 9300 bit strings for each $I_\mathrm{dc}$, which are obtained by the pattern recognition method described in Supplementary Note 5 and Supplementary Figure 5.


In summary, we have demonstrated experimentally that the origin of complexity in chaotic NCVOs is controlled by switching of the vortex core, leading to unpredictable sequences of three distinct oscillatory waveform patterns. The chaos was characterised by testing the sensitivity to initial conditions and correlation-dimension analysis of the time-resolved magnetoresistance signal of the NCVO. In the incommensurate state, the NCVO switches chaotically between two different patterns inherited from the neighboring commensurate dynamics. We show that these pattern sequences can be reduced, upon proper partitioning of reconstructed phase space, to a bit stream, interpreted here as the NCVO's symbolic dynamics, which coincides with core-switching events.  The complexity of the bit-sequence generated at 100 MHz rates shows that NCVOs can achieve close-too maximum entropy when they are at the centre of their incommensurate states. This property can be used to design a true random number generator at the nanoscale. In addition, knowing the underlying structure of the temporal chaotic dynamics and its connection to the timing of the core-switching events, one could design experimental strategies to control electrically the core dynamics to encode information secretly. This paves the way for nanoscale chaos-based information processing using the nonlinear dynamics of spintronic devices.

\section*{Methods}

\subsection*{Sample fabrication and measurements}
The nanocontacts are fabricated using the atomic force microscope nano-indentation method~\cite{Bouzehouane:2003kc} on the top of the sputtered deposited multilayer with the composition $\mathrm{SiO_2}$ /Cu (40 nm)/Co (20 nm)/Cu (10 nm)/$\mathrm{Ni_{81}Fe_{19}}$ (20 nm)/Au (6 nm)/photoresist (50 nm)/Au (nanocontact)~\cite{Devolder:2019uo}. The diameter of the contact is $\sim 20$ nm. The vortex is first nucleated by reversing the free layer magnetization with an in-plane applied magnetic field in the presence of a $I_\mathrm{dc}$ = 16 mA current applied through the nanocontact. The vortex gyration around the nanocontact results in magnetoresistance oscillations that are detected after amplification as voltage fluctuations in the frequency domain by a spectrum analyser and in the time domain by a single-shot oscilloscope. RF switches are used to connect either of these two equipments to the sample, hence allowing for both time- and frequency- domain measurements to be made sequentially. The experiments are conducted in a cryostat at liquid nitrogen temperature (77 K) to better isolate the chaotic dynamics, which is a deterministic process but can appear as athermal noise, from thermal fluctuations which are true stochastic processes. Further details of the experimental setup and measurement procedure are described elsewhere~\cite{PetitWatelot:2012be, Devolder:2019uo}.

\subsection*{Thermal noise filtering from time traces}
To improve the signal-to-noise ratio of the experimental time traces, we used an averaging filter. We collected similar short-term waveforms ($\sim 7.5$ ns) from full time series by calculating convolutions, then averaged over them. This method is applicable in our system because the output time traces are composed of only two or three patterns even in the chaotic regime. The details are given in Supplementary Note 3 and Supplementary Figure 3.

\subsection*{Time-delay embedding and phase-space reconstruction}
A time-delay embedding procedure is used to form an $m$-dimensional vector space related to the original phase space by a diffeomorphism preserving topological invariants of the original attractor, if $m > 2d_\mathrm{A}$ with $d_\mathrm{A}$ dimension of the original attractor~\cite{Takens:81}. The vectors in the reconstructed phase-space are obtained from univariate time-resolved series as follows
\begin{equation}
	\mathbf{v}_n^{(m)} = [V(t_n), V(t_n-\tau),\dots,V(t_n-(m-1)\tau)],
\end{equation}
with the measured voltage, $V(t_n)$, sampled at discrete $t_n = n\Delta t$ with $\Delta t = 12.5$ ps the experimental sampling period. In this study, we choose the time-delay embedding $\tau \approx 1/(4f_0)$ and the embedding dimension $m = 3$.

\subsection*{Poincar{\'e} section and symbolic analysis}
To simplify the definition of the Poincar{\'e} section ($x_3 = 0$) in the symbolic analysis, we apply a unitary transformation to the lag coordinates (which does not affect the topological equivalence between the reconstructed and original phase spaces) and form $\mathbf{x}_n^{(m)} = \mathbb{U}\mathbf{v}_n^{(m)}$, where $\mathbb{U}$ is a rotation matrix,
\begin{equation}
	\mathbb{U} =
	\begin{bmatrix}
		1 & 0 & 0 \\
		0 & \cos{\theta_1} & -\sin{\theta_1} \\
		0 & \sin{\theta_1} & \cos{\theta_1}
	\end{bmatrix}
	\begin{bmatrix}
		\cos{\theta_3} & -\sin{\theta_3} & 0 \\
		\sin{\theta_3} & \cos{\theta_3} & 0 \\
		0 & 0 & 1
	\end{bmatrix},
\end{equation}
with $\theta_1 = -20$\si{\degree} and $\theta_3 = 67$\si{\degree}. The partition on the Poincar{\'e} section is set as $x_2 = 0.76 x_1 - 0.005$ (Fig.~\ref{Fig:Symbolic}a-c) delimiting two regions, $R_\mathrm{A}$ and $R_\mathrm{B}$. This choice, despite being arbitrary, allows us to capture both the incommensurate and commensurate regimes of the NCVO ($I_\mathrm{dc} = 12.6 - 14.0$ mA). A reconstructed orbit is encoded with symbolic sequences, whose length is determined by the number of times the reconstructed attractor intersects the two regions.

\subsection*{Computation of correlation dimension}
The correlation dimension $D_c$ provides insight on the fractal dimension of an attractor and hence is used for the detection of chaos from the filtered time-series. Its computation relies on the Grassberger-Procacia (GP) algorithm~\cite{GP:83a,GP:83b} involving the correlation sum $C(m,\epsilon)$, which gives the average number of neighbouring vectors $\mathbf{x}_j^{(m)}$ within the range $\epsilon>0$ from any given vectors $\mathbf{x}_i^{(m)}$ of the attractor obtained from the time-delay embedding procedure. It is defined as
\begin{equation}
	C(m,\epsilon){=}\frac{2}{(N{-}n_{T})(N{-}1{-}n_{T})}\sum_{i{=}1}^N\sum_{j{=}i{+}1{+}n_{T}}^N \Theta\left(\|\mathbf{x}_i^{(m)}{-}\mathbf{x}_j^{(m)}\|{-}\epsilon\right),
\end{equation}
with $\epsilon$ representing the typical radius of the neighbourhood surrounding the vector $\mathbf{x}_i^{(m)}$, $\|{\cdot}\|$ the norm-2, and $\Theta$ the Heaviside function. To avoid the bias induced by finite-size effects of the time series and time-correlation of neighbouring vectors, we introduce the Theiler condition $|i-j|>n_{T}$ to select eligible neighbours for the computation. The presence of self-similarity imposes the correlation sum to approximately satisfy a linear growth in log-scale and hence as a constant value (plateau) for its derivative. Hence, the correlation dimension $D_c$ is given by
\begin{equation}
	D_\mathrm{c} = \lim_{\epsilon\to 0}\lim_{N\to\infty} \frac{\partial \log C(m,\epsilon) }{\partial \log \epsilon}.
\end{equation}
In our analysis, we have used this approach after normalizing the experimental filtered time series as described in the previous section. We use $N = 1.2 \times 10^5$ samples and $n_{T} = 15$ for the Theiler condition, with the embedding dimension $m$ chosen between $6$ and $10$, and the neighborhood radii $\epsilon$ in the range $[10^{-2},10]$.

\subsection*{Computation of the Shannon block entropy}
We consider a random source of $n$-bit words from the dictionary $\{s_i\}_{n_s}$ with $1 \le n_s \le 2^{n}$. The words are obtained by sliding a window of $n$ bits in width along the bit stream of length $N_b$ resulting from the symbolic analysis of the NCVO dynamics. The mathematical definition of the Shannon block entropy of the $n$-bit word source is given by
\begin{equation}
	H_n = -\sum_{s_i\in\{s_1,\dots,s_{n_s}\}} p(s_i)\log p(s_i),
\end{equation}
with $p(s_i)$ the probability of appearance of symbol $s_i$. We can then determine the entropy per binary symbol (or entropy rate with a maximum value at 1 bit/symbol) using the limit $h = \lim_{n\to\infty} H_n/n$.
In finite binary sequences of length $N_\mathrm{b}$, the use of blocks of length $n$ usually leads to more robust estimates of the entropy compared with the direct estimation from the bit sequence. The probability of each word is determined with the likelihood estimator $\hat{p}(s_i) \approx \#(s_i)/N_s$ with $Ns = N_b-n+1$. Finally, we use the upper limit for the block size given by
\begin{equation}
N_s h \ge n \, 2^{n} \log{2} 
\end{equation}
for binary words as suggested in Ref.~\onlinecite{Pezard:09}. Due to the finite size $N_\mathrm{b} = 9300$ bits for the bit stream obtained from the NCVO's symbolic dynamics, we use block lengths in the range $n\in\{3,6 \}$.

\subsection*{Computation of the Lempel-Ziv complexity}
The Lempel-Ziv complexity $C_\mathrm{LZ}$ of a binary sequence measures the number of patterns present and is the basis of  LZ77 compression~\cite{LZ:76,LZ:77}. For a large binary sequence of size $n$, it can be shown that the number of patterns $c(n)$ behaves asymptotically like the ratio $n/\log_2 n$. In a sequence of length $n\gg 1$, we can use this ratio as a normalization factor for the number of patterns in order to ensure that the complexity measure remains bounded in the range $C_\mathrm{LZ}\in[0,1]$. In order to compute $C_\mathrm{LZ}$, we use the algorithmic procedure presented in Ref.~\onlinecite{Kaspar:87}. If the binary sequence is generated by a stationary and ergodic process, the Lempel-Ziv complexity coincides with the entropy rate $h$ \cite{Pezard:09} in the limit of large $n$. Here, the complexity $C_\mathrm{LZ}$ is computed from a sequence of $N_\mathrm{b} = 9300$ bits.

\subsection*{Data availability}
The data sets generated and/or analysed during the current study are available from the corresponding author on reasonable request.

\section*{References}
\bibliography{reference}{}

\section*{Acknowledgements}
The authors thank Cyrile Deranlot and St{\'e}phanie Girod for their assistance in film growth and sample preparation. This work was supported by the Agence Nationale de la Recherche under Contract No. ANR-17-CE24- 0008 (CHIPMuNCS), the Horizon2020 Research Framework Programme of the European Commission under Contract No. 751344 (CHAOSPIN), and the French RENATECH network. The Chaire Photonique is funded by the European Union (FEDER), Ministry of Higher Education and Research (FNADT), Moselle Department, Grand Est Region, Metz Metropole, AIRBUS-GDI Simulation, CentraleSup{\'e}lec, and Fondation CentraleSup{\'e}lec.

\section*{Author contributions}
J-V. K., S.P.-W., and V.C. designed the study. K.B. and V.C. fabricated the samples. T.D. designed and implemented the experimental set-up. M-W.Y., J.L., and T.D. performed the high-frequency electrical measurements. M-W.Y., D.R., and J-V.K. analyzed and interpreted the data. M-W.Y. performed the simulations and interpreted the results. M-W.Y. and D.R. prepared the manuscript. All authors edited and commented on the manuscript.

\section*{Competing interests}
The authors declare no competing interests.

\section*{Additional Information}

\textbf{Supplementary information} is available for this paper.

\textbf{Correspondence and requests for materials} should be addressed to M.-W.Y.

\newpage
\renewcommand{\figurename}{Supplementary Figure}
\renewcommand{\theequation}{S\arabic{equation}}

\makeatletter
\renewcommand*{\fnum@figure}{{\normalfont\bfseries \figurename~\thefigure}}
\renewcommand*{\@caption@fignum@sep}{\textbf{ $|$ }}
\makeatother

\setcounter{figure}{0}
\setcounter{equation}{0}

\section*{Supplementary Figures}

\begin{figure}[ht]
\centering\includegraphics[width = 8.5cm]{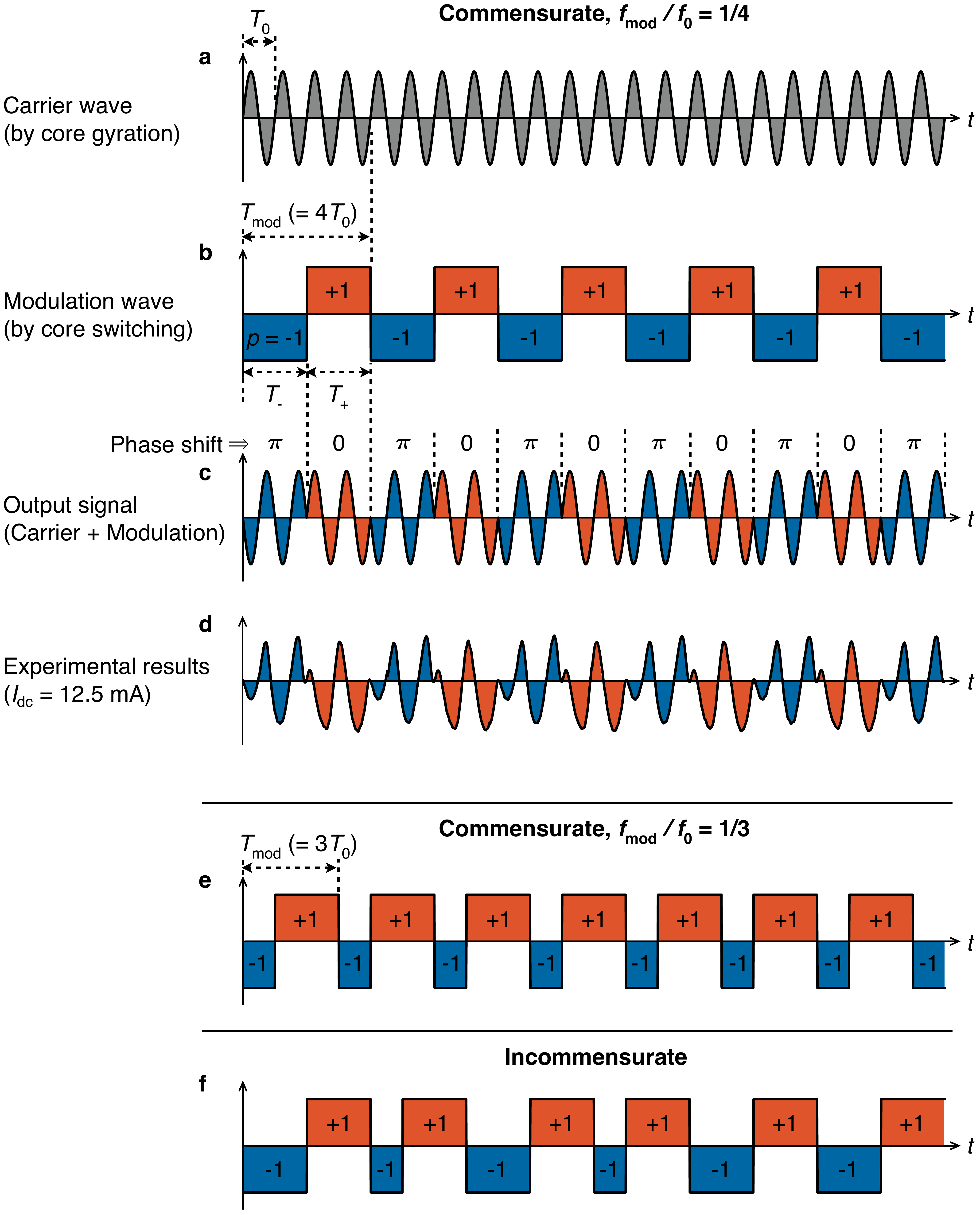} 
\caption{\label{Fig:Comm}
\textbf{Ratio between fundamental and modulation frequencies.} \textbf{a}, A sinusoidal carrier wave with a period of $T_0 = 1/f_0$. \textbf{b}, A square modulation wave for the case of $f_\mathrm{mod} / f_0 = 1/4$. $T_{+}$ and $T_{-}$ are duration times for $p$ = $+1$ and $-1$, respectively.The red and blue colours represent the core polarity, $p$ = $+1$ and $-1$, respectively. \textbf{c}, The output signal obtained from the carrier and modulation by a phase shift. The shifted phase are represented above the graph. The red and blue colours correspond to those in b. \textbf{d}, An experimentally measured time-resolved voltage oscillation at the commensurate state ($f_\mathrm{mod} / f_0 = 1/4$). \textbf{e}, as in b but in the case of $f_\mathrm{mod} / f_0 = 1/3$. \textbf{f}, as in b but in the incommensurate case.
   }
\end{figure}

\begin{figure*}[ht]
	\centering
	\includegraphics[width = 12cm]{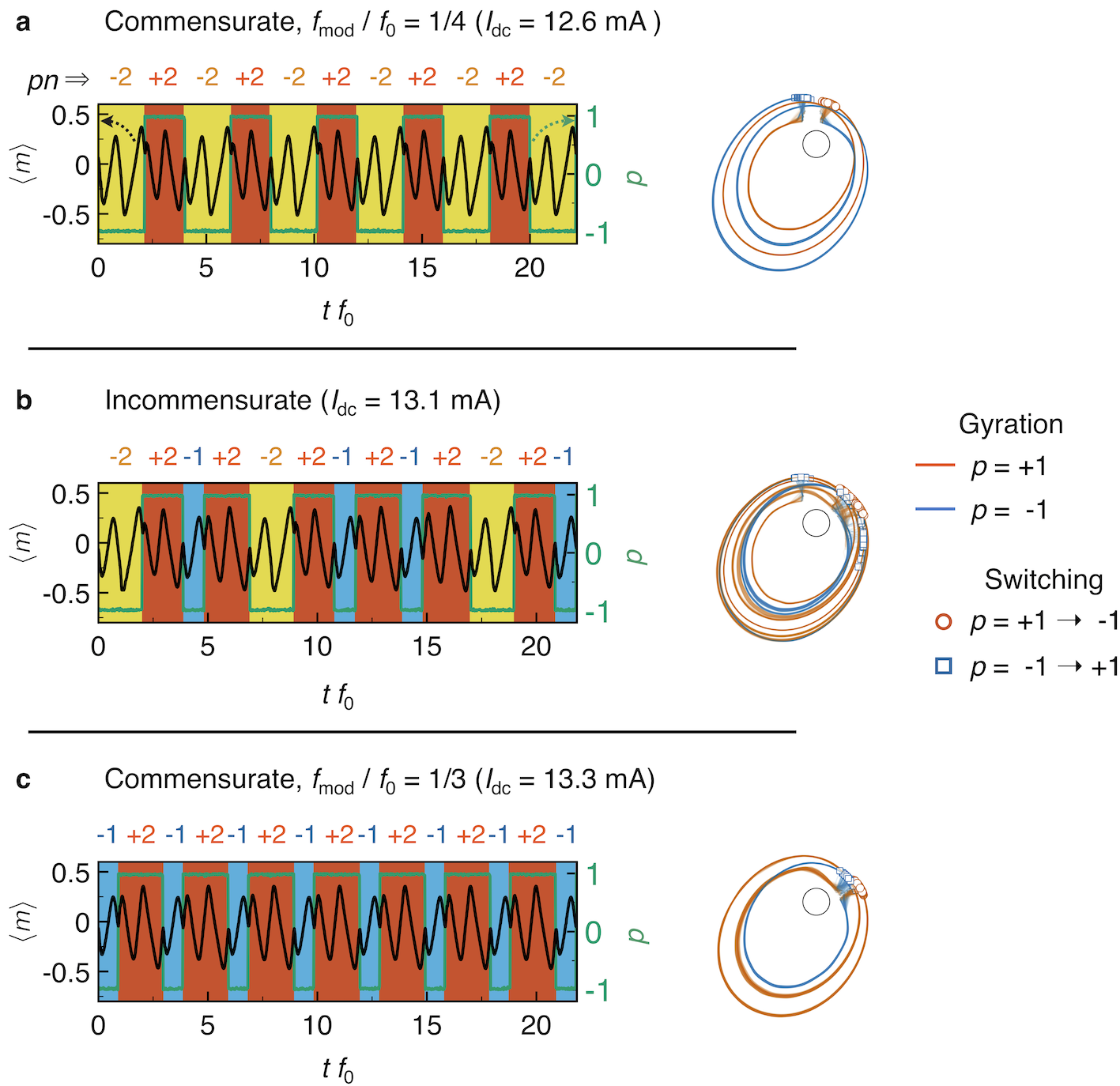}
	\caption{\label{Fig:Pattern}
	\textbf{Pattern generation from micromagnetic simulations.} \textbf{a}, Time evolution of an average of magnetization, $\langle m \rangle$ for the commensurate ($f_\mathrm{mod} / f_0 = 1/4$) (black line). $x$ axis is normalised by $1/f_0$, where $f_0$ is the gyration frequency. The green line is the core polarisation associated with the right axes. The colour regions indicate corresponding $pn$ patterns defined in Fig.2g in the main text. On the right side, the corresponding core trajectories are presented. The red and blue lines are the core trajectories when $p$ = $+1$ and $-1$, respectively. The red circles and blue squares indicate the core switching positions from $p$ = $+1$ to $-1$ and from $-1$ to $+1$, respectively. \textbf{b}, as in a but in the incommensurate case. \textbf{c}, as in a but in the case of $f_\mathrm{mod} / f_0 = 1/3$.
	}
\end{figure*}

\begin{figure*}[ht]
	\centering
	\includegraphics[width = 14cm]{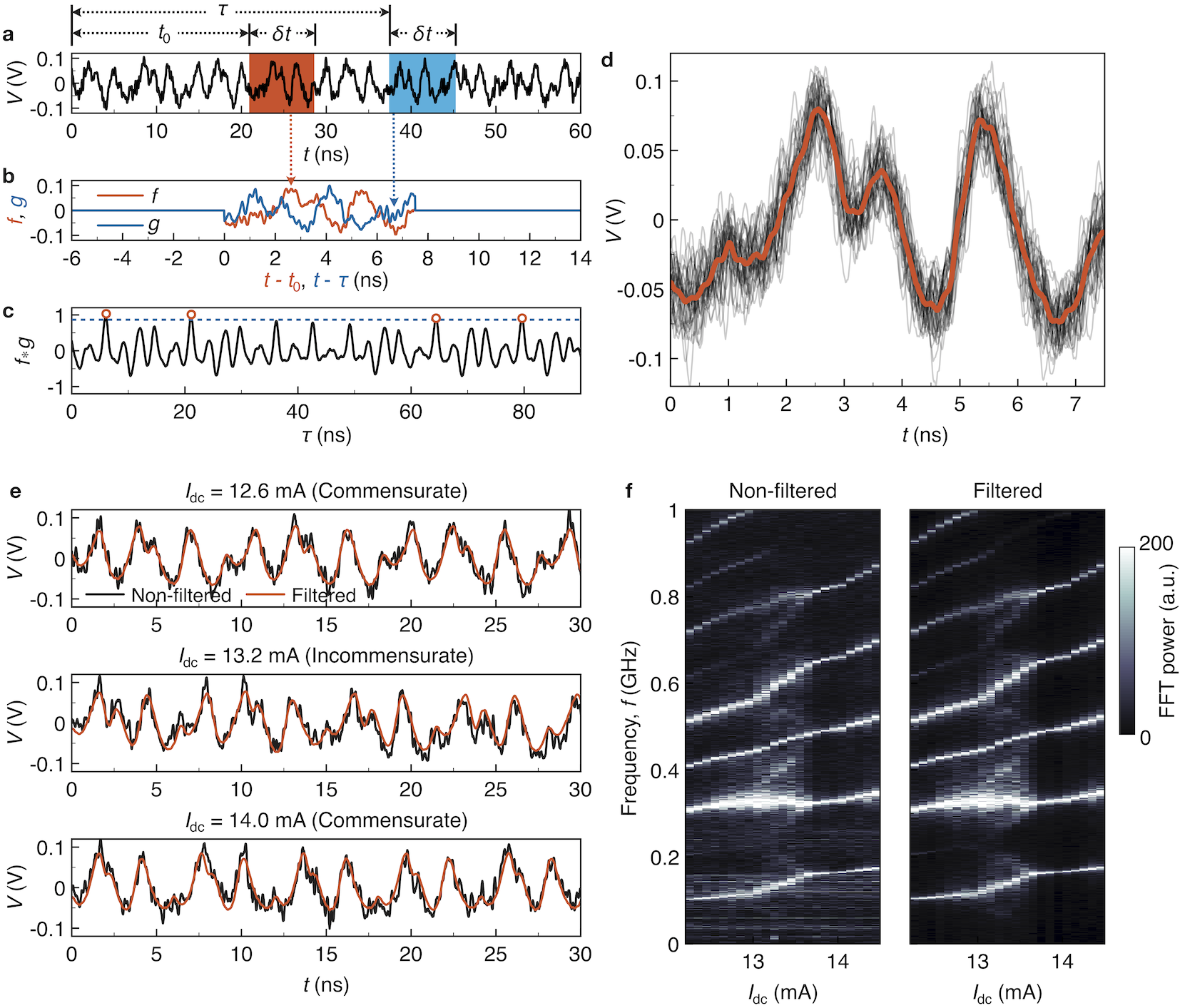}
	\caption{\label{Fig:Filter}
	\textbf{Noise reduction filter.} \textbf{a}, A measured non-filtered time trace at $I_\mathrm{dc}$ = 13.2 mA. The red and blue areas represent $t_0 \leq t < t_0+\delta t$ and $\tau \leq t < \tau+\delta t$ which are the regions of the functions of $f$ and $g$, respectively. \textbf{b}, $f(t - t_0)$ (red line) and $g(t - \tau)$ (blue line) which are correspond with the colour regions in a. \textbf{c}, $(f(t - t_0) * g(t - \tau))$ as a function of $\tau$. The dashed line represents a tolerance value and red circles indicate the valid peak positions ($\tau_i = \tau_1, \tau_2, \cdots$) which heights are larger than the tolerance. \textbf{d}, Collected short-term waveforms from $\tau_i$, $V(\tau_i \leq t \leq \tau_i + \delta t)$ (black lines). The red curve is a filtered data obtained by averaging the black lines. \textbf{e}, Measured non-filtered signals (black) and filtered signals (red). \textbf{f}, Power of the fast Fourier transformation (FFT) as a function of the input current, $I_\mathrm{dc}$, calculated from the non-filtered (left) and filtered (right) time traces.
	}
\end{figure*}

\begin{figure*}[ht]
	\centering
	\includegraphics[width = 10cm]{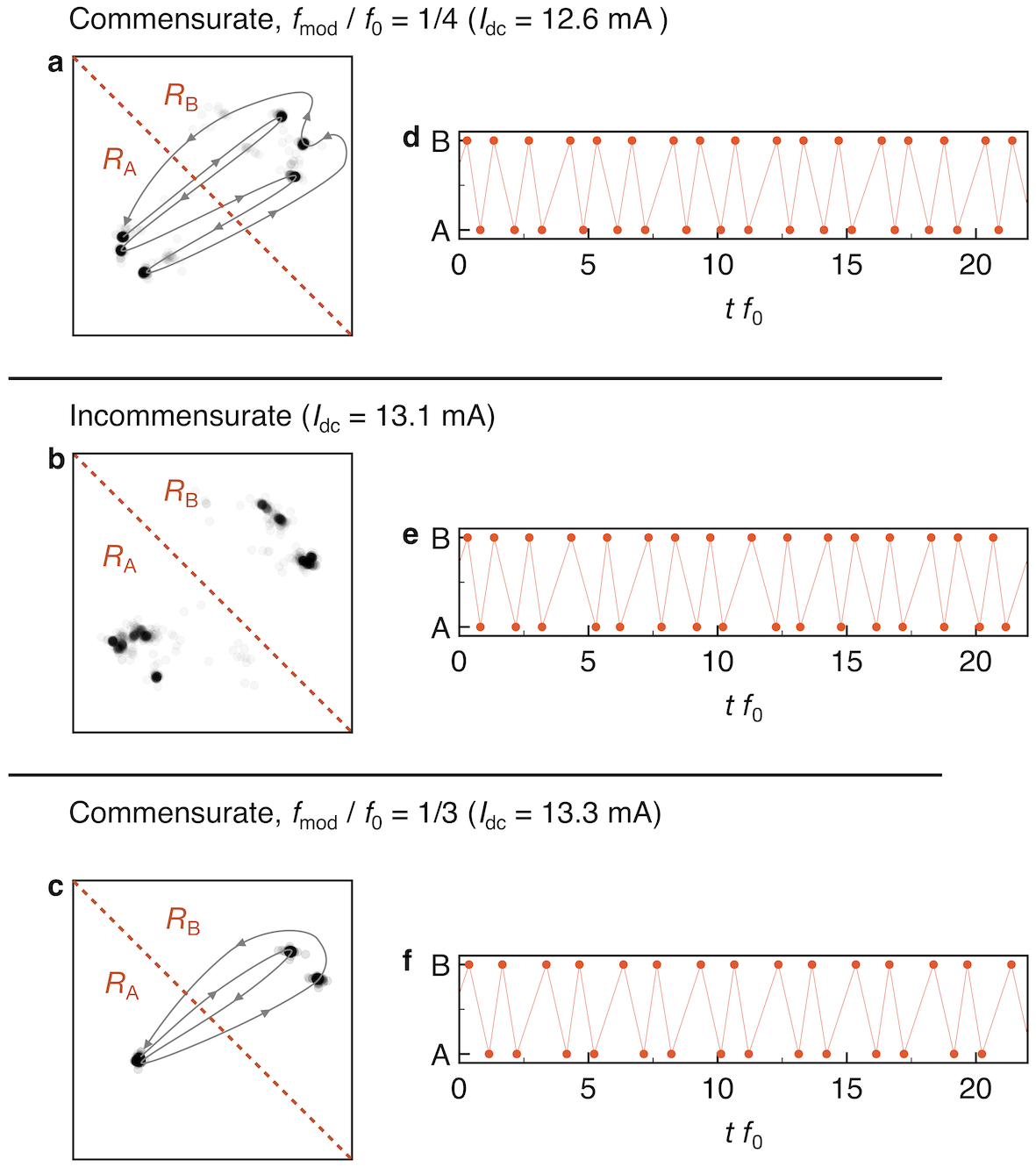}
	\caption{\label{Fig:Attractor}
	\textbf{Symbolic dynamics using a different partition.} \textbf{a-c}, Poincar{\'e} maps for $I_\mathrm{dc}$ = 12.6 mA, 13.2 mA, and 14.0 mA. The red dashed line is a new partition to divide the regions, $R_\mathrm{A}$ and $R_\mathrm{B}$ for symbolic dynamics. The gray arrows in a and c show a dynamics of the intersection points in the case of the commensurate state. \textbf{d-f}, Symbolic dynamics based on the newly defined partition.
	}
\end{figure*}

\begin{figure*}[ht]
	\centering
	\includegraphics[width = 14cm]{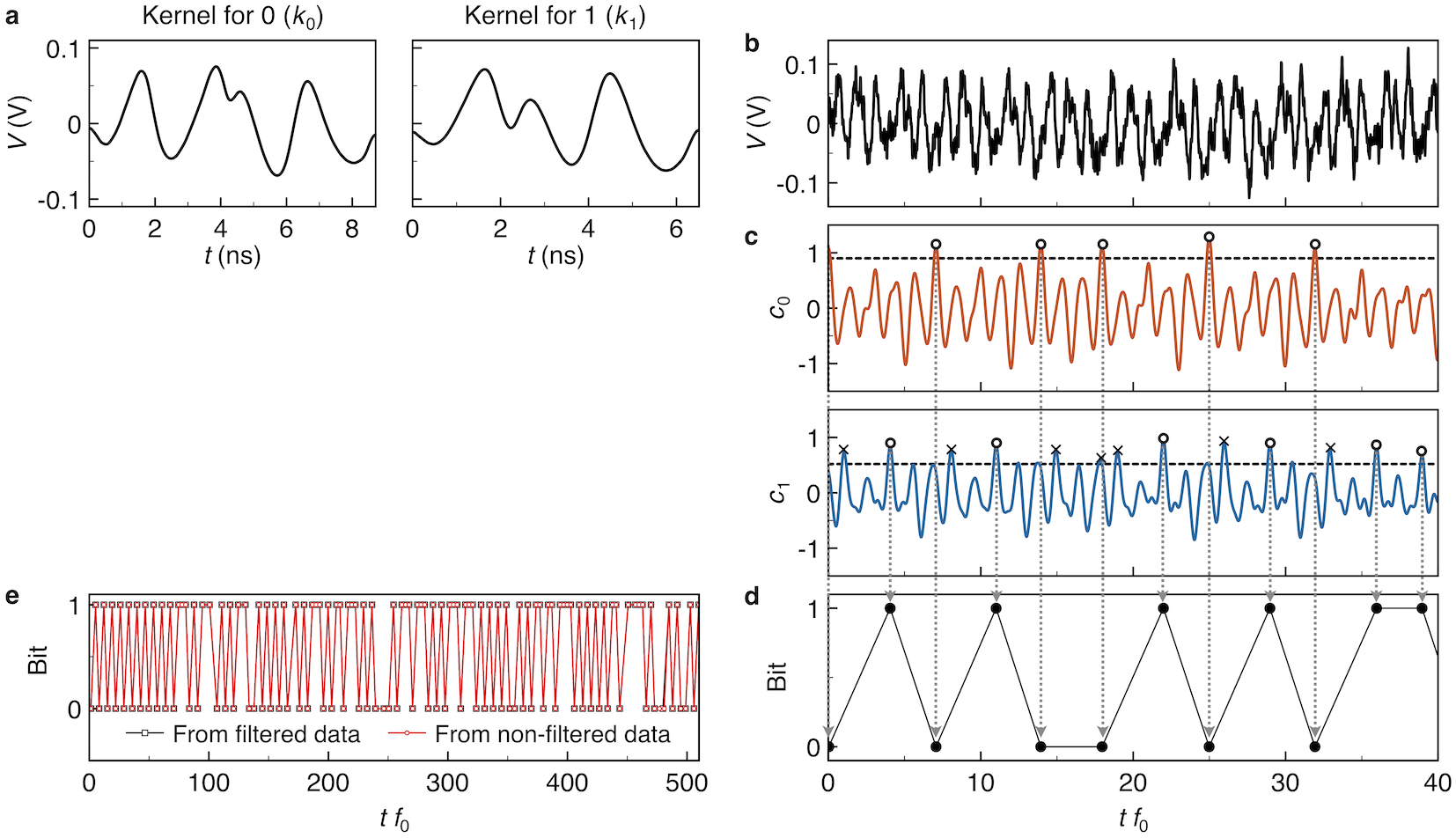}
	\caption{\label{Fig:Recognition}
	\textbf{Bit sequences from non-filtered data by convolutions.} \textbf{a}, Kernel functions of ``0'' and ``1'' for $I_\mathrm{dc}$ = 13.2 mA. \textbf{b}, A non-filtered time trace at $I_\mathrm{dc}$ = 13.2 mA. \textbf{c}, Convolutions, $c_0$ and $c_1$, between the non-filtered time trace and the kernels, $k_0$ (Top) and $k_1$ (Bottom), respectively. The dashed lines show prescribed tolerances for $c_0$ and $c_1$. The open circles are valid peaks for embedding bits. Cross marks in $c_1$ are fallacious peaks which need to be ignored . \textbf{d}, An obtained bit sequence recognised from the non-filtered data by the convolutions. The gray dashed arrows from c show the corresponding peaks. \textbf{e}, Red circles indicate bit sequences directly obtained from the non-filtered data, and black squares represent the bits calculated from the filtered data.
	}
\end{figure*}

\begin{figure*}[ht]
\centering\includegraphics[width = 14cm]{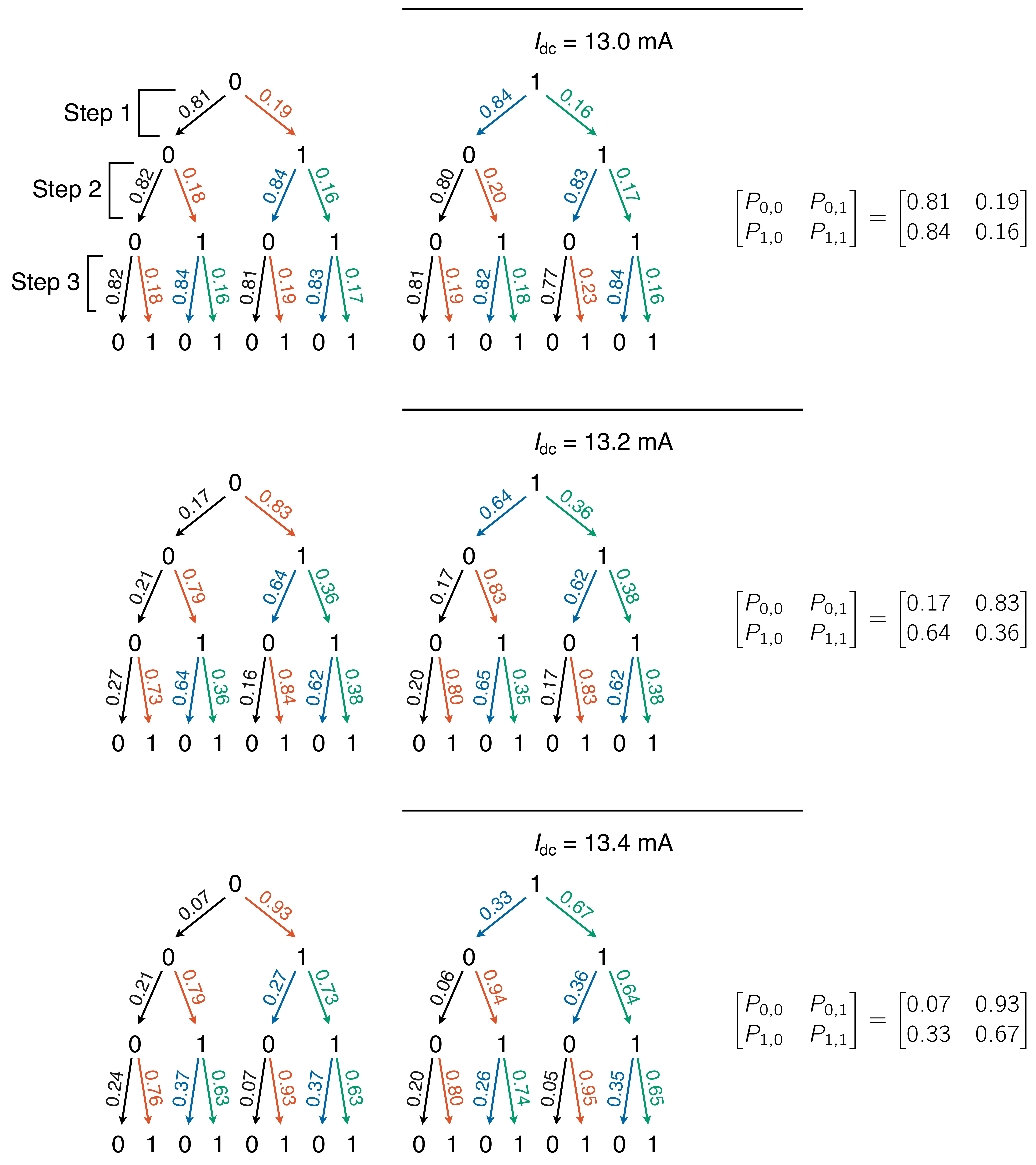}
\caption{\label{Fig:Tree}
\textbf{Tree diagrams and Markov matrices.} (Left) Tree diagrams of moving for three steps. Numbers at the arrows show probabilities of the moving including their previous states. The black, red, blue, and green colours represent the movings of 0 $\rightarrow$ 0, 0 $\rightarrow$ 1, 1 $\rightarrow$ 0, and 1 $\rightarrow$ 1, respectively. (Right) Markov matrices for each $I_\mathrm{dc}$ by assuming one step memory dynamics. $P_{i, j}$ indicates the probability of the moving from $i$ to $j$.
   }
\end{figure*}
\clearpage
\newpage

\section*{Supplementary Notes}

\subsection{Ratio between fundamental and modulation frequencies}
In this note we discuss the meaning of the frequency ratio, $f_\mathrm{mod} / f_0$, as explained in the main text (Fig. 1c), where $f_0$ and $f_\mathrm{mod}$ are fundamental and modulation frequencies, respectively. The oscillatory signals from the nanocontact vortex oscillator (NCVO) can be considered as phase-modulated sinusoidal waves by the vortex core switching, because the core reversal inverts a sense of the core gyration. The sinusoidal oscillations of the carrier wave originates from the vortex core gyration (Supplementary Figure~\ref{Fig:Comm}a), whose frequency corresponds with the gyration frequency, $f_0 = 1/T_0$, where $T_0$ is a time period of the gyration; the core switching generates the a square-shaped phase modulation signal of which frequency is $f_\mathrm{mod} = 1/(T_\mathrm{mod}) = 1/(T_{+}+T_{-})$, where $T_\mathrm{mod}$ is a period of the modulation and $T_{+}$ and $T_{-}$ are dwell times for the upward core ($p = +1$) and downward core ($p = -1$), respectively. $p$ is a core polarity. An example of the modulation of $f_\mathrm{mod} = f_0/4$ and the modulated output signal are represented in Supplementary Figure~\ref{Fig:Comm}b and c, respectively. The schematic output signal is qualitatively in a good agreement with the experimental results (compare with Supplementary Figure~\ref{Fig:Comm}d).

From Supplementary Figure~\ref{Fig:Comm}a-c, we can conclude that $f_\mathrm{mod} / f_0$ can be expressed approximately as
\begin{equation}\label{Eqn:fmodf0}
	\frac{f_\mathrm{mod}}{f_0} = \frac{T_0}{T_\mathrm{mod}}  = \frac{1}{n_{+}+n_{-}},
\end{equation}
in the typical commensurate state, where $n_{+} = T_{+} f_0$ and $n_{-} = T_{-} f_0$ are required gyration numbers for the core switching in the cases of $p$ = $+1$ and $-1$, respectively. In Supplementary Figure~\ref{Fig:Comm}b, for example, $n_{+}$ = $2$ and $n_{-}$ = $2$, thus $f_\mathrm{mod} / f_0 = 1/4$. In Supplementary Figure~\ref{Fig:Comm}e, $n_{+}$ = $2$ and $n_{-}$ = $1$, thus $f_\mathrm{mod} / f_0 = 1/3$. For the incommensurate case, we replace the gyration numbers by their time-averaged values, $\langle n_{+} \rangle$ and $\langle n_{-} \rangle$, respectively.

\subsection{Chaotic pattern generation using micromagnetic simulations}
In the main text (Fig. 2a-c), we show chaotic sequences of patterns, $pn$, from the experiments, where $p$ and $n$ are a core polarity and a required gyration number for the core switching, respectively. Here we show identical pattern generations using micromagnetic simulations at zero temperature.

For the micromagnetic simulations, we solve the Landau-Lifshitz-Gilbert equation with in-plane spin-torque terms using the MuMax3 software~\cite{Vansteenkiste:2014et}. We chose a $1280 \times 1280 \times 20$ \si{nm^3} film which is uniformly discretized with $512 \times 512 \times 1$ finite difference cells. The magnetic parameters used here correspond to those of a NiFe film: an exchange stiffness constant of $A_\mathrm{ex} = 10$ pJ/m, a saturation magnetization of $M_\mathrm{s} = 0.8$ MA/m, a Gilbert damping constant of $\alpha = 0.013$, and a spin polarisation of $P = 0.5$. The spatial distributions of the current and its associated Oersted field were computed using a finite-element method~\cite{PetitWatelot:2012iz}. The initial state used for the simulations is obtained by sweeping an in-plain magnetic field with an applied dc current of 10 mA. To mimic an asymmetry between the $p = +1$ and $-1$, we apply both a perpendicular dc field of $B_z = -18$ mT and an in-plane DC field of $B_y = 3$ mT simultaneously. In the experiments, the asymmetry is induced by an exchange interaction between the vortex and its magnetic configurations around the vortex.

The simulations were conducted for the commensurate state (Supplementary Figure~\ref{Fig:Pattern}a and \ref{Fig:Pattern}c) and the incommensurate state (Supplementary Figure~\ref{Fig:Pattern}b), then we extracted a spatial average of magnetization, $\langle m \rangle$, in the saturated magnetisation direction of the reference layer, which is proportional to the output voltage amplitude. We assume that the magnetisation of the reference layer is saturated to \ang{-135} from the $x$ axis, and calculated $\langle m \rangle$ only in 100 nm from the center of the nanocontact.

The results are plotted in Supplementary Figure~\ref{Fig:Pattern} for the commensurate and incommensurate states as well as the corresponding core trajectories. The obtained $\langle m \rangle$ are qualitatively in good agreements with the experimental results (see Fig.~2a-c in the main text); they show sinusoidal oscillations delimited by cusp points. We also plot the time evolution of the core polarity, $p$ (green lines in Supplementary Figure~\ref{Fig:Pattern}), which directly shows that the cusp positions correspond with the core reversal as discussed in the main text.

In the commensurate cases (Supplementary Figure~\ref{Fig:Pattern}a and \ref{Fig:Pattern}c), we can find periodic repetition of two patterns of [$-2$, $+2$] and [$-1$, $+2$], respectively. On the other hand, in the incommensurate state (Supplementary Figure~\ref{Fig:Pattern}b), the sequence is composed of three patterns, [$-2$, $-1$, $+2$], and they are erratically ordered. This result is in a good agreement with the experimental result in Fig.~2a-c in the main text, and shows that the chaotic signals at the incommensurate state have deterministic characteristics, although thermal noise affects the dynamics in real experiments.

\subsection{Noise reduction filtering}
To reduce thermal noise from the measured time traces, we refine them by averaging short-period waveforms. This method is available in this case, since the output signals are composed by only few simple patterns even in the incommensurate state (Supplementary Figure~\ref{Fig:Pattern}), therefore, we can find sufficient numbers of similar short-term waveforms from long-term measured data. The process are stated below.

\begin{enumerate}
\item $V(t)$ is the measured time-domain signal. Choose an interesting short-time period ($t_0 \leq t < t_0 + \delta t$), then obtain the kernel function, $f(t - t_0) = V(t - t_0) \chi (t)$, where $\chi (t)$ is a step function defined as
\begin{equation}
	\begin{aligned}
		\chi (t) =
		\begin{cases}
			1 & \text{if } 0 \leq t < \delta t, \\
			0 & \text{if otherwise}.
		\end{cases}
	\end{aligned}
\end{equation}
In this study, $\delta t = 7.5$ ns (red region in Supplementary Figure~\ref{Fig:Filter}a and red curve in Supplementary Figure~\ref{Fig:Filter}b).
\item Set $g(t - \tau) = V(t - \tau) \chi (t)$  (blue region in Supplementary Figure~\ref{Fig:Filter}a and blue curve in Supplementary Figure~\ref{Fig:Filter}b).
\item Calculate the convolution $f(t - t_0) * g(t - \tau)$ as a function of $\tau$ (Supplementary Figure~\ref{Fig:Filter}c).
\item Find positions of peaks, $\tau_i$, whose heights are larger than a predetermined tolerance value (Supplementary Figure~\ref{Fig:Filter}c).
\item Collect the waveforms from $V(\tau_i \leq t < \tau_i + \delta t)$  (black curves in Supplementary Figure~\ref{Fig:Filter}d).
\item Average all the collected short-term traces (red curve in Supplementary Figure~\ref{Fig:Filter}d).
\item Repeat the same process for different $t_0$.
\end{enumerate}

In both the time and frequency domains, we compare the filtered signal with the non-filtered data (Supplementary Figure~\ref{Fig:Filter}e and \ref{Fig:Filter}f), which show reasonably similar results, although some high-frequency information can be lost by the filtering through the averaging process.

\subsection{Partition on Poincar{\'e} map for symbolic dynamics}
For symbolic dynamics, we arbitrarily choose a proper Poincar{\'e} surface of section and partition. The surface and partition in the main text are one of a successful case for generating meaningful bit sequences. Here, by using an unsuitable partition on the same surface, we show an example of failing to generate bit sequences. In Supplementary Figure~\ref{Fig:Attractor}a-c, we display the Poincar{\'e} maps on the Poincar{\'e} section and their dynamics using gray arrows. The red dashed line is the new partition which divide clearly the maps. However, as shown in Supplementary Figure~\ref{Fig:Attractor}d-f, the embedded symbols with the partition always generate only periodic sequences of [A,B] even in the incommensurate state. Please compare the symbolic dynamics with the dynamics generated in the main text (see Fig.~3d-f). This result shows that the new partition is not suitable to be used for generating meaningful bit sequences.

\subsection{Pattern recognition}
In Figs. 3d-f in the main text, we show generated bit sequences from filtered time series, however, the data filtering takes substantial time. Here we discuss a method to directly extract the bits from the non-filtered data by using kernel functions. The process is described below: 
\begin{enumerate}
\item Prepare kernel functions for 0 and 1, $k_0$ and $k_1$, (Supplementary Figure~\ref{Fig:Recognition}a) from preliminary measurements.
\item Measure a new time trace (Supplementary Figure~\ref{Fig:Recognition}b), then calculate the convolutions, $c_0$ (for $k_0$) and $c_1$ (for $k_1$), between the non-filtered data and the kernels as a function of shifting time (Supplementary Figure~\ref{Fig:Recognition}c).
\item Find the peaks from $c_0$ and $c_1$ whose heights are higher than tolerance values (circles and cross marks in Supplementary Figure~\ref{Fig:Recognition}c). Record the corresponding bit when the peaks appear (Supplementary Figure~\ref{Fig:Recognition}d).
\end{enumerate}
We discard spurious peaks in $c_1$ which are very close to the peak positions in $c_0$. The cross marks in the bottom panel of Supplementary Figure~\ref{Fig:Recognition}c indicate the spurious peaks. We do not use these peaks for recording ``1''. In Supplementary Figure~\ref{Fig:Recognition}e, we plot the bit sequence obtained using the convolutions (red dots and curve), which is in a good agreement with that obtained from the filtered data (black squares and lines).

\subsection{Tree diagrams}
From bit sequences in the incommensurate cases, we obtain tree diagrams (Supplementary Figure~\ref{Fig:Tree}). The result shows dependences of the moving probability on their history which is related to the grammar of the symbolic dynamics. If the probabilities do not depend on their history, the dynamics can be considered suitable for a Markov system.

For the trees, we use a finite length of bit sequences ($\sim$ 9300), thus it is not easy to be assured whether the dynamics is suitable for the Markov system or not. In the case of $I_\mathrm{dc}$ = 13.0 mA and 13.2 mA, all probabilities does not vary largely like one step memory dynamics. However, when $I_\mathrm{dc}$ = 13.4 mA, $P_{0,0}$ is varied from 0.05 to 0.24 by the history. This change of the probability may originate from the dependence on the previous steps or may come from a lack of a number of the samples, which can increase the error of the probability.

In the main text, we assume one step memory dynamics, and obtained the probabilities, $P_{i,j}$ based on the Markov system. We also represent Markov matrices for each $I_\mathrm{dc}$ which corresponds with Fig.~4d in the main text.

\end{document}